\documentclass[twocolumn,aps,prl,groupedaddress]{revtex4-1}

\usepackage{graphicx}
\usepackage{dcolumn}
\usepackage{bm}
\usepackage{amsmath,amssymb,amsfonts,bbm}
\usepackage{color}
\usepackage{braket}

\begin{document}

\title{Nodal Arc in Disordered Dirac Fermions: Connection to Non-Hermitian Band Theory}
\author{Micha{\l} Papaj}
\author{Hiroki Isobe}
\author{Liang Fu}
\affiliation{Department of Physics, Massachusetts Institute of Technology, Cambridge, Massachusetts 02139, USA}

\begin{abstract}

We show that Dirac fermion systems in two dimensions generally exhibit disorder-induced nodal arc replacing the nodal point and tilted Dirac cone, provided that the two components of the Dirac fermion correspond to two distinct orbitals unrelated by symmetry. This result is explicitly demonstrated using renormalization group analysis in a disordered Dirac model that we introduce, where the disorder potential acts differently on the two orbitals. As we show by numerical simulations and self-consistent Born approximation calculation, this drives the system into a new strongly disordered phase.

\end{abstract}


\maketitle

Disordered Dirac fermions in two dimensions have been studied for decades in a variety of contexts \cite{Fradkin, Guruswamy,review}, including the integer quantum Hall transition \cite{Ludwig} and  disordered unconventional superconductivity  \cite{Nersesyan,SenthilFisher,Bocquet,Altland,Foster}.
The interest to this field has been reinforced since the discovery of graphene and topological insulators  \cite{McCann,Koshino,Ostrovsky,Aleiner,Altland2,Ostrovsky2,Nomura,Bardarson,Ryu2,Noro, Mong, FuKane}. 
In general, massless Dirac fermions in solids arise from various types of internal degrees of freedom, such as electron spin in topological insulator surface states, the two sublattices in graphene, or particle/hole excitations in cuprates.
In these three cases, the two components of Dirac fermions, i.e., the two degenerate states at the Dirac point, belong to  a symmetry doublet associated with time-reversal, spatial inversion and particle-hole symmetry respectively.

A second kind of massless Dirac fermions exists in solids, whose two components correspond to distinct degrees of freedom unrelated by any symmetry. For example, Dirac fermions on the (001) surface of topological crystalline insulators SnTe and Pb$_{1-x}$Sn$_x$Se  are comprised of the cation Sn/Pb orbital and the anion Te/Se orbital \cite{ZeljkovicMappingunconventionalorbital2014}. In heavy fermion semimetals, Dirac fermions can emerge from the hybridization of $f$ and $d$ bands \cite{KoziiNonHermitianTopologicalTheory2017, Dai}, which have very different masses. 

In the presence of disorder, the two distinct orbitals, from which the Dirac fermion is formed, are generally expected to have different scattering rates. Then, the self-energy of the Dirac fermion in the disorder averaged single-particle Green's function acquires an {\it orbital-dependent} imaginary part. The existence of two scattering rates---a generic property of Dirac fermion of the second kind---unnoticed in previous studies, has an important consequence that is only recognized very recently. As shown by one of the authors \cite{KoziiNonHermitianTopologicalTheory2017}, in such case the imaginary part of self-energy not only broadens the energy spectrum, but also alters the energy-momentum dispersion. It transforms the original Dirac point into a ``nodal arc''---a line of band degeneracy without fine tuning. The two ends of this nodal arc are exceptional points, where the inverse of the Green's function becomes non-diagonalizable.  
This bulk nodal arc connecting a pair of exceptional points is topologically robust and unique to non-Hermitian band theory recently developed for finite-lifetime quasiparticles  \cite{ShenTopologicalBandTheory2017}. 
It is also shown that finite-lifetime effects lead to a flat band or a nodal line segment in type-II Weyl semimetals in three dimensions \cite{WeylFlatBand}.

In this Letter, we report the finding of a nodal arc and a new universality class in disordered two-dimensional (2D) Dirac fermions of the second kind, comprised of two distinct orbitals that do not belong to any symmetry doublet.
Our model includes a random potential that acts on the two orbitals differently. This type of random potential, not considered in the standard treatment of disordered Dirac fermions,
appears naturally in real materials (see below).
Our renormalization group (RG) analysis shows the disorder is marginally relevant, driving the system into a strongly disordered phase.
The disorder reduces the quasiparticle weights of the two orbitals at low energy by different amounts.
As a result, it generates a tilt of the Dirac cone, even when it is initially absent.  In the disordered phase, the two orbitals acquire different lifetimes, which inevitably leads to a bulk nodal arc replacing the Dirac point in the clean limit. 
The nodal arcs and the tilt of Dirac cones are also observed directly in our numerical simulations.

We consider a 2D Dirac Hamiltonian
\begin{equation}
\label{eq:model}
H_0(\bm{r}) = \psi^\dagger(\bm{r}) [-i(v_x \sigma_z - w \sigma_0)\partial_x -iv_y \sigma_x\partial_y ] \psi (\bm{r}),
\end{equation}
and the disorder of the form
\begin{equation}
H_\text{dis}(\bm{r}) = V(\bm{r}) \psi^\dagger(\bm{r}) \eta \psi(\bm{r}),
\end{equation}
where $\psi=(\psi_1, \psi_2)^T$ is a two-component fermion field, $\sigma_i$ ($i=x,y,z$) are the Pauli matrices, $\eta$ is a $2\times 2$ Hermitian matrix depending on the type of disorder to be specified below, and $V(\bm{r})$ is a random function.
The velocity parameter $w$ describes the tilt of the Dirac cone along the $x$ direction in the absence of disorder, where the velocity of the steep and gentle sides of the cone are $v_x\pm w(>0)$ depending on the sign of $w$.  The velocity along the $y$ direction is given by $v_y$.  We restrict the tilt to satisfy $w^2<v_x^2$, so that the Fermi surface remains closed. Such massless and tilted Dirac fermions described by $H_0$ appear in a number of materials, including (001) surface states of SnTe \cite{Ando, SodemannFu} and organic conductor $\alpha$-(BEDT-TTF)$_2$I$_3$ \cite{tilted1, tilted2, Isobe}.

We assume that the random function $V(\bm{r})$ is spatially uncorrelated and obeys a Gaussian distribution $P[V] \propto \exp [-\int d^2r V^2(\bm{r})/(2\Delta)]$. The random function is characterized by
$
\langle V (\bm{r}) V (\bm{r}')\rangle = \Delta \delta(\bm{r}-\bm{r}'),
$
where $\Delta$ $(>0)$ characterizes the strength of disorder.

The $2\times 2$ matrix $\eta$ specifies the type of disorder potential.  Previous studies of disordered Dirac systems considered the cases where the orbitals forming Dirac fermions are related by a symmetry.  
In real materials where the two orbitals are unrelated by a symmetry, however, disorder strengths for the two orbitals are allowed to be different; for example, $\eta$ may have a form $\eta=A\sigma_0+B\sigma_z$ with two independent constants $A$ and $B$.
Such a form is excluded when the two orbitals are related by a symmetry since the symmetry is restored after disorder averaging.

For simplicity, we henceforth consider the limit $\eta=\eta_{11}\equiv(\sigma_0+\sigma_z)/2$, where only the $\sigma_z=+1$ orbital is disordered.
This highly asymmetric case is relevant for the surface state of topological crystalline insulator Pb$_{1-x}$Sn$_x$Te, where the two components $\sigma_z= \pm 1$ of the surface Dirac fermion correspond to Sn/Pb and Se orbitals respectively. Since the Pb sites are substituted with Sn, the disorder potential on the $\sigma_z=+1$ orbital is naturally much stronger.

In the following, we will show that the disorder characterized by $\eta_{11}$ ($A/B=1$) is marginally relevant in the RG sense.  We also confirmed that the disorder is marginally relevant and our conclusion remains valid for a wide range of the ratio $A/B$.
The analysis for general cases will be reported elsewhere \cite{general}.

We study disordered-averaged electron spectral function and density of states, which can be computed from the disorder-averaged Green's function: $\bar{G}(\omega) = \langle G(\omega) \rangle \equiv [\omega-H_0-\Sigma(\omega)]^{-1}$, where $G = (\omega -H_0-H_\text{dis})^{-1}$ is the one-particle Green's function before disorder average, and $\Sigma$ is the self-energy---a non-Hermitian $2\times 2$ matrix.
Note that the translational invariance is statistically recovered after the disorder average. The poles of $\bar{G}$ on the complex plane determine the quasiparticle energy spectrum and lifetime in the presence of disorder.

We use the replica method to evaluate disorder averages. With the replica method, we take the disorder average and obtain the Euclidean action
\begin{align}
S =& \sum_a \int d\tau d^2r \psi^\dagger_a [ \partial_\tau -i(v_x \sigma_z - w \sigma_0)\partial_x -iv_y \sigma_x\partial_y] \psi_a \notag\\
&-\sum_{ab} \frac{\Delta}{2} \int d\tau d\tau' d^2r (\psi^\dagger_a\eta\psi_a)(\tau) (\psi^\dagger_b\eta\psi_b)(\tau'),
\end{align}
where $a,b(=1,\ldots , n)$ are replica indices. The quartic term, effectively working as an interaction between replicas, is generated by the disorder average.


\begin{figure}
\centering
\includegraphics[width=\hsize]{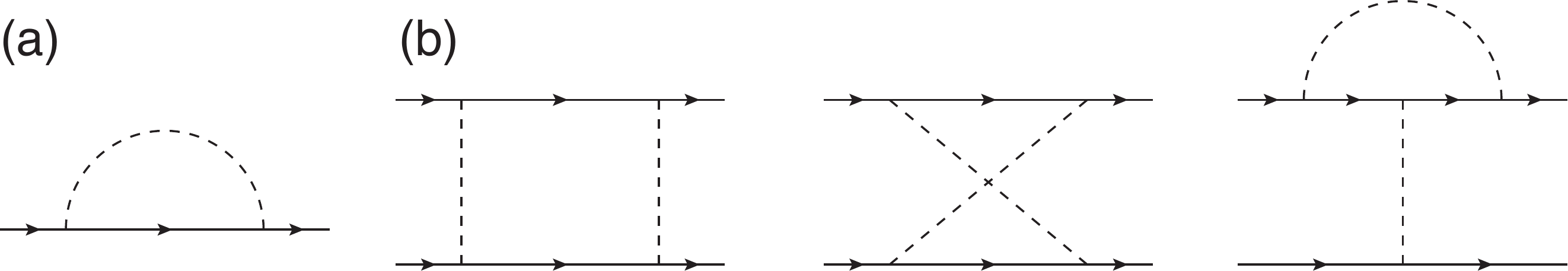}
\caption{
One-loop diagrams that contribute to the RG equations.
Solid and dashed lines correspond to the fermion field and disorder, respectively.
The two-point diagram (a) contributes to the field renormalization, and the four-point diagrams (b) give corrections to the disorder strength.
}
\label{fig:diagram}
\end{figure}

Now we examine whether the parameters are relevant or irrelevant, by a perturbative RG calculation to one-loop order (Fig.~\ref{fig:diagram}).  The calculations give the scale dependence of parameters, arising from resummations of logarithmic divergences.

The two-leg diagram [Fig.~\ref{fig:diagram}(a)] gives the renormalization of quasiparticle weight $Z=(1-\operatorname{Re}\partial\Sigma/\partial\omega)^{-1}$.  We have two different quasiparticle weights $Z_{11}$ and $Z_{22}$ for the two orbitals $\sigma_z=\pm 1$, respectively.  Here, the self-energy $\Sigma(\omega)$ depends only on frequency $\omega$, but not on momentum. Therefore, the renormalization of the velocity parameters $v_x$, $v_y$, and $w$ owes solely to the renormalization of the quasiparticle weights.  The four-leg diagrams [Fig.~\ref{fig:diagram}(b)] bear disorder strength renormalization.
After evaluating the one-loop diagrams, we obtain the following set of RG equations:
\begin{gather}
\frac{dv_x}{dl} = -\alpha v_x, \quad
\frac{dv_y}{dl} = -\alpha\frac{v_x}{v_x-w} v_y, \quad
\frac{dw}{dl} = \alpha v_x , \notag\\
\frac{d\Delta}{dl} = 4\alpha \frac{w}{v_x-w}\Delta, \notag\\
\frac{dZ_{11}}{dl} = -\frac{\alpha}{2}\frac{v_x}{v_x-w}\Delta, \quad
\frac{dZ_{22}}{dl} = 0,
\end{gather}
where $\alpha = \Delta/[4\pi\sqrt{(v_x^2-w^2)v_y^2}]$ is the dimensionless quantity, $l=\ln(\Lambda/\epsilon)$, $\Lambda$ is the UV energy cutoff, and $\epsilon$ is the energy scale of interest.

Even when there is initially no tilt of the Dirac cone, the tilt $w(>0)$ is generated by the disorder. This is seen from the decreasing quasiparticle weight of the $\sigma_z=+1$ orbital as the energy scale goes down.  For $k_y=0$, this orbital corresponds to the energy branch of $(v_x-w)k_x$.  In low energies, the decreasing $Z_{11}$, accompanied by the change of $\operatorname{Re}\Sigma(\omega)$, reduces $(v_x-w)$.
In contrast, since the $\sigma_z=-1$ orbital is free from disorder, $Z_{22}$ remains constant and so does $(v_x+w)$.
The tilt $w>0$ increases the density of states of the $\sigma_z=+1$ orbital, which helps the disorder strength $\Delta$ grow as the energy approaches the Dirac point.  Thus, $\Delta$ is marginally relevant, driving the system into a disordered phase.
In the course of increasing $\Delta$, the tilt keeps growing whereas the two velocity parameters $v_x$ and $v_y$ decrease.  Those flows stop as $w\to v_x$, where the gentle slope of the Dirac cone becomes flat and the increasing density of states further drives the flow to the strongly disordered phase.

To study the properties of the disordered phase, the self-energy $\Sigma$ is evaluated by the self-consistent Born approximation.  Within this approximation, the self-consistent equation becomes
\begin{equation}
\label{eq:self-consistent}
\Sigma(\omega) = \Delta \int'\frac{d^2k}{(2\pi)^2} \eta \bar{G}(\bm{k},\omega) \eta.
\end{equation}
$\int'$ indicates an integration with the cutoff $\Lambda$. 
In the present model with $\eta=\eta_{11}$, the self-energy $\Sigma(\omega)$ is nonzero only for the $\sigma_z=+1$ orbital; $\Sigma(\omega)=\Sigma_{11}(\omega)\eta_{11}$.
The self-energy is calculated for the two regions $|\epsilon|\ll|\Sigma|$ and $|\epsilon|\gg|\Sigma|$, which are separated at the energy scale
\begin{equation}
\Gamma_0 = 2\Lambda \sqrt{\frac{v_x-w}{v_x+w}} \exp\left( -\frac{v_x-w}{v_x+w} \frac{1}{\alpha} \right).
\end{equation}
$\Gamma_0$ corresponds to the energy scale where the one-loop RG breaks down. Close to $\Gamma_0$, the velocity $(v_x-w)$ approaches zero, which destroys perturbative expansion with respect to $\alpha$.

The retarded self-energy $\Sigma_{11}^R(\omega)$ is obtained from Eq.~\eqref{eq:self-consistent} in a series of $\omega$ for $|\omega|\ll\Gamma_0$, and in a series of $\alpha$ by iterations for $|\omega|\gg\Gamma_0$:
\begin{align}
&\Sigma_{11}^{R}(\omega)  \notag\\
=&
\begin{cases}
-\dfrac{2v_x(v_x-w)}{(v_x+w)^2} \dfrac{\omega}{\alpha} - i\Gamma_0 & |\omega|\ll\Gamma_0, \\
-\alpha \dfrac{v_x}{v_x-w} \left[ \omega \ln \left(\dfrac{(v_x^2-w^2) \Lambda^2}{v_x^2 \omega^2}\right) + i\pi |\omega| \right] & |\omega|\gg\Gamma_0.
\end{cases}
\end{align}

Importantly, it shows that the Green's function has a finite imaginary part even at $\omega=0$, appearing only in the $\sigma_z=+1$ orbital.
The energy spectrum of the quasiparticles is obtained as zeros of $\bar{G}$, i.e., $\det [E-H_0(\bm{k})-\Sigma(E)]=0$.
The quasiparticle energy dispersion becomes now complex-valued because of the non-Hermitian component of $\Sigma$. Furthermore, the parameters of the Hamiltonian are renormalized by the reduced quasiparticle weight $Z_{11}$, giving the eigenvalues for $|E| \ll \Gamma_0$ as 
\begin{equation}
E_\pm=\frac{(\tilde{v}_2- v_1)k_x-i\tilde{\Gamma}_0}{2}\pm\sqrt{\left[\frac{(\tilde{v}_2+v_1)k_x-i \tilde{\Gamma}_0}{2}\right]^2+v_y^2k_y^2}
\end{equation}
with $v_1 = v_x + w$, $\tilde{v}_2 = Z_{11} (v_x - w)$ and $\tilde{\Gamma}_0=Z_{11}\Gamma_0$.

In the clean limit, two linearly dispersing energy bands touch at a Dirac point.
With disorder, the quasiparticle energy obtains an imaginary component, and the real parts of the two energy branches coalesce along the line $|k_y|\leq \tilde{\Gamma}_0/(2v_y)$ ($k_x=0$).
Here, we can observe the formation of a nodal arc in a disordered Dirac model, which lies along the $y$ direction and terminates at exceptional points $(0,\pm\tilde{\Gamma}_0/(2v_y))$.

A nodal arc is absent when the two orbitals are related by a symmetry and disorder is characterized by $\eta=\sigma_0$ or $\sigma_i$.  In such cases, the energy dispersion is simply smeared by the same finite lifetime for the two orbitals because of the symmetry. The self-consistent equation \eqref{eq:self-consistent} also confirms that the self-energy is proportional to $\sigma_0$.

It is important to point out that the formation of the nodal arc by disorder is observed even at zero temperature $T=0$, because plane wave (or Bloch wave) states as quasiparticles are not eigenstates due to the lack of translational symmetry by disorder.  Therefore, unlike the arc due to interactions, a nodal arc formed by disorder is independent from thermal effects, which contribute largely to thermal broadening of energy dispersions.

\begin{figure*}
\centering
\includegraphics[width=\hsize]{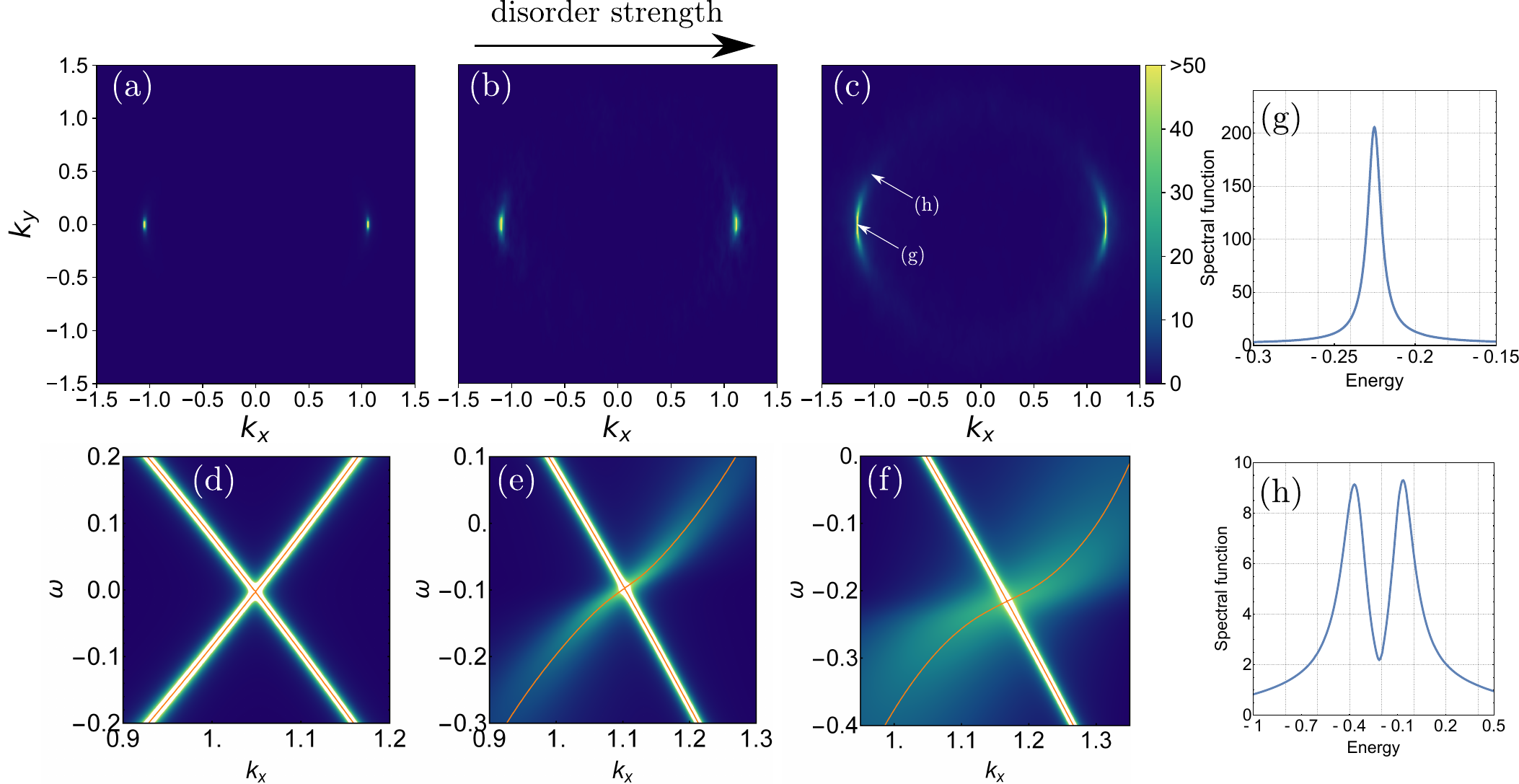}
\caption{(a)--(c) Spectral function $A(\bm{k}, \omega)$ slices across the relevant fragment of the Brillouin zone for $\omega$ corresponding to the open contour Fermi surface, calculated for $t=1$, $t_y=0.25$, $u=3.0$ and disorder strengths $V_0 = 0.5, 3.0, 4.5$, respectively. As the disorder strength increases, the nodal arc becomes longer. (d)--(f) Spectral function $A(\bm{k}, \omega)$ slices for $k_y=0$ and the same parameters as respective panels above. As the disorder increases, the Dirac cone tilts. (g), (h) Energy dependence of the spectral function for a single point in $\bm{k}$ space for a point in the middle of the arc and a point on the faint contour away from the nodal arc, chosen as indicated by white arrows on panel (c).
}
\label{fig:TB_SCBA}
\end{figure*}

\begin{figure}
\centering
\begin{tabular}{cc}
\includegraphics[width=\hsize]{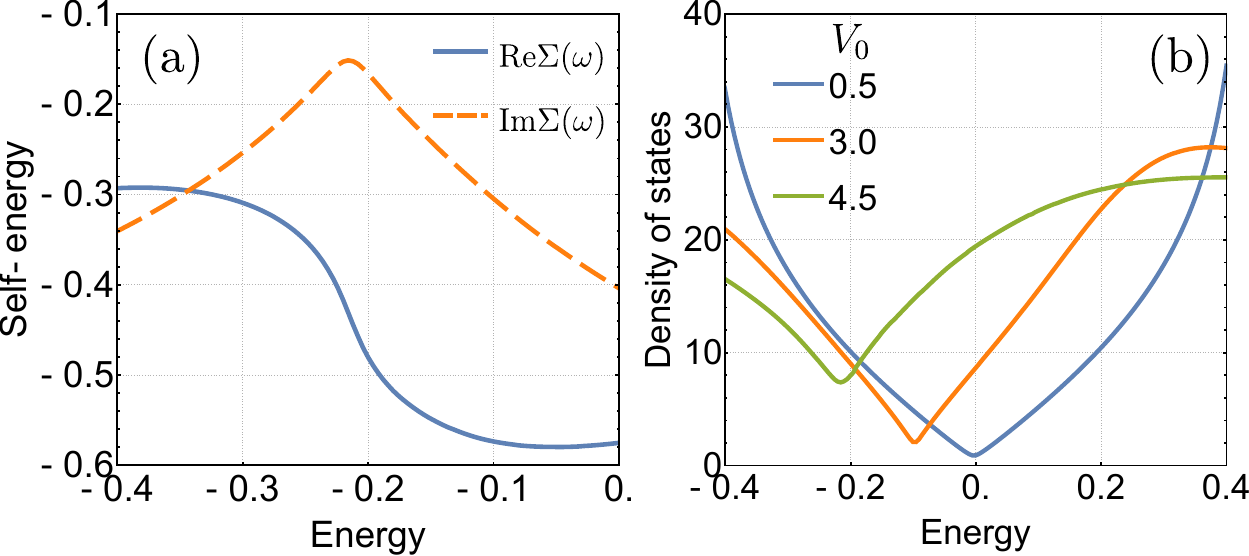}
\end{tabular}
\caption{(a) The self-energy $\Sigma(\omega)$ calculated using self-consistent Born approximation with parameters corresponding to Fig.~\ref{fig:TB_SCBA}(c). (b) Total integrated density of states $\rho(\omega)$. The position of the minimum, which shifts to lower energies correspond to the position of the Dirac points in the case of infinite lifetimes.
}
\label{fig:self_energy_dos}
\end{figure}

Next, we perform numerical simulations on a square lattice with the periodic boundary conditions to confirm the observations above.
Because of the fermion doubling problem, the lattice model necessarily has a paired Dirac cone in the Brillouin zone. The tight-binding Hamiltonian that we use is
\begin{gather}
\hat{H}_0 = -t\sum_{\langle ij \rangle} c_i^\dagger \sigma_z c_j + t_y \sum_{i} (ic_{i+\hat{y}}^\dagger \sigma_x c_i + \text{H.c.}) + u \sum_i c^\dagger_i \sigma_z c_i, \notag\\
\hat{H}_\text{dis} = \sum_i V_i c^\dagger_i \eta_{11} c_i.
\end{gather}
For $0<u<4t$, $\hat{H}_0$ has two Dirac points at $k_x a=\pm\arccos[(u-2t)/(2t)]$, $k_ya=0$, where $a$ is a lattice constant. We assume that there is no tilt of the Dirac cones in the clean limit. By a series expansion near $k_x a = \arccos[(u-2t)/(2t)]>0$ point, we can establish the connection to the Dirac model from Eq.~\eqref{eq:model} by $v_x=\mathrm{sgn}(u-2t) \sqrt{(4t-u)u}$, $v_y=-2 t_y$, $w=0$.
For the purpose of numerical simulations, we choose a random potential $V_i$ obeying a uniform distribution over an interval $[-V_0/2, V_0/2]$, and its spatial correlation function is $\langle V(\bm{r}_i) V(\bm{r}_j)  \rangle = \frac{V_0^2}{12} \delta (\bm{r}_i-\bm{r}_j)$. In all the simulations the lattice consists of 800$\times$800 sites. We choose $t=1$, $t_y=0.25$, $u=3$, $a=1$ and change the disorder strength to observe the evolution of nodal arcs. We use MUMPS and Kwant \cite{MUMPS, KWANT} packages for the numerical calculations.

For a particular disorder distribution, we can compute the retarded Green's function $G^R(\bm{r},\bm{r}',\omega) = (\omega + i\delta - H_0-H_\text{dis})^{-1}(\bm{r},\bm{r}',\omega)$, where $\delta$ is a small positive quantity.  The disorder average is taken by computing $G^R$ for 2000 independent disorder realizations. (A result for a single disorder realization is presented in the Supplemental Material (SM) \cite{SM}.)  We can then reinstate translational symmetry in the averaged quantity $\bar{G}^R(\bm{r}-\bm{r}',\omega) = \langle G^R(\bm{r},\bm{r}',\omega ) \rangle$, which allows the Fourier transformation $\bar{G}^R(\bm{k},\omega)$. The spectral function is $A(\bm{k}, \omega) = - \frac{1}{\pi} \operatorname{Im} \operatorname{Tr} \bar{G}^R(\bm{k},\omega)$.

Figs.~\ref{fig:TB_SCBA}(a)--(c) show the spectral function $A(\bm{k},\omega)$ at the energy where the two bands touch, with the disorder strength $V_0=0.5$, 3.0, 4.5, respectively.
We can observe the nodal arcs extending to the $y$ direction, with stronger disorder yielding longer nodal arcs.
The curvature of the nodal arcs reflects the energy dispersion of the tight-binding model $\hat{H}_0$, which corresponds to a higher-order effect with respect to $k$ in the linearized model, Eq.~\eqref{eq:model}.

The disorder-averaged Green's function can be alternatively obtained from the self-consistent equation \eqref{eq:self-consistent} in successive iterations of numerical integrations. Now we replace the parameter $\Delta$ with $V_0^2/12$ and the integration is now over the Brillouin zone.
Using the self-energy obtained by the self-consistent calculation, we similarly obtain spectral functions, which are in excellent agreement with those in Figs.~\ref{fig:TB_SCBA}(a)--(c); see SM \cite{SM}. This supports the conclusions that the self-energy is largely momentum independent and the important effects are due to its energy dependence instead. We also obtain the slices of $A(\bm{k}, \omega)$ in the $k_y=0$ plane [Figs.~\ref{fig:TB_SCBA}(d)--(f)], with solid lines indicating the position of the poles of $\bar{G}^R(\bm{k},\omega)$.
We confirm that the tilt of Dirac cones is generated by the disorder, even when it is initially absent. Since only the orbital with a gentle slope of the energy dispersion is affected by the disorder, the spectral function is smeared by $\operatorname{Im}\Sigma$, and its peaks are smaller due to the decreasing quasiparticle weight $Z$.

We also check that the contour we observe in Fig.~\ref{fig:TB_SCBA}(c) is indeed a nodal arc by determining the spectral function as a function of energy for particular $\bm{k}$ values on and off the arc as indicated by the white arrows. In Figs.~\ref{fig:TB_SCBA}(g) and (h), we observe a single peak in the spectral function for the point on the contour and two peaks on the faint structure outside of the arc.

The self-energy $\Sigma(\omega)$ obtained from the self-consistent Born approximation is shown in Fig.~\ref{fig:self_energy_dos}(a).
Note that $\operatorname{Re}\Sigma_{11}(\omega)$ shifts the band crossing point to $\omega-\operatorname{Re}\Sigma_{11}(\omega)/2=0$, since $\Sigma(\omega)=\Sigma_{11}(\omega) (\sigma_0+\sigma_z)/2$. Crucially, close to the band touching point, $\operatorname{Re}\Sigma_{11}$ is linearly dependent on energy and the slope increases with disorder strength causing the tilting of the Dirac cone, which is consistent with result of RG analysis.
Moreover, at the band touching point, $\operatorname{Im}\Sigma$ has a dip, but it remains finite, supporting the existence of the nodal arcs. This result agrees with the solution of the self-consistent equation for the linearized model.
The shift of the band touching point is also evident in the density of states
$\rho(\omega) = \int_\text{BZ} d^2k A(\bm{k}, \omega)$, which is depicted in Fig.~\ref{fig:self_energy_dos}(b). The minimum of $\rho(\omega)$ moves to lower energy values with larger disorder strength. The minima match the solutions of $\omega-\operatorname{Re}\Sigma_{11}(\omega)/2=0$ (see Fig.~S1 in SM \cite{SM}).

In summary, we studied a disordered 2D Dirac system with two orbitals not belonging to a symmetry doublet, and reported a nodal arc and a new universality class in disordered 2D Dirac fermions. By the RG analysis we have shown that asymmetric disorder for the two orbitals is marginally relevant, and in the strongly disordered phase, the inequality of the quasiparticle weights and lifetimes of the two orbitals yields a tilt of the Dirac cone and a bulk nodal arc, replacing the Dirac point. The nodal arc is formed by the non-Hermitian effect of the self-energy generated by the disorder, which is present even at $T=0$. This conclusion is supported by numerical calculations using a tight-binding model and self-consistent Born approximation, which show appearance of a nodal arc in the spectral function.


\textit{Acknowledgment}. This work was supported by  the David and Lucile Packard Foundation. 

M.P. and H.I. contributed equally to this work.

\newpage
\onecolumngrid
\setcounter{equation}{0}
\setcounter{figure}{0}
\setcounter{page}{1}
\renewcommand{\thefigure}{S\arabic{figure}}
\renewcommand{\theequation}{S\arabic{equation}}

\begin{center}
\textbf{\large Supplemental Material for ``Nodal Arc in Disordered Dirac Fermions: Connection to Non-Hermitian Band Theory''}
\end{center}

\section{Additional details on the numerical simulation}

The numerical simulation has been performed on an $800 \times 800$ tight-binding lattice with periodic boundary conditions described by the following Hamiltonian 
\begin{equation}
\label{eq:lattice_ham}
\hat{H}_0 = -t\sum_{\langle ij \rangle} c_i^\dagger \sigma_z c_j + t_y \sum_{i} (ic_{i+\hat{y}}^\dagger \sigma_x c_i + \text{H.c.}) + u \sum_i c^\dagger_i \sigma_z c_i . 
\end{equation}
In momentum space, this corresponds to
\begin{equation}
\label{eq:momentum_ham}
\tilde{H}_0(\mathbf{k}) = -2 t( \cos(k_x a) + \cos(k_y a)) \sigma_z + t_y \sin(k_y a) \sigma_x + u \sigma_z. 
\end{equation}
We can now calculate the spectral function in two alternative ways:  performing direct calculation of the spectral function on the lattice and determining the self-energy using the self-consistent Born approximation.

In the first case, we model the Anderson type disorder in the sample by adding random potential $U(\mathbf{r}, \mathbf{r}')$ to the single particle Hamiltonian $\hat{H}_0$. Because we want to study the impact of asymmetric scattering of the two orbitals in the model, to simplify the analysis we take $U(\mathbf{r}, \mathbf{r}')$ as an onsite term on just a single orbital:
\begin{equation}
U(\mathbf{r},\mathbf{r}') = \delta_{\mathbf{r}\mathbf{r}'}V(\mathbf{r}) \begin{pmatrix} 1 & 0 \\ 0 & 0\end{pmatrix}. 
\end{equation}
In all the simulations we are choosing $V(\mathbf{r})$ from a uniform distribution of random numbers that belong to the range $[-V_0/2, V_0/2]$, where $V_0$ is the disorder strength. Such a random potential satisfies the conditions
\begin{equation}
\label{eq:disorder_properties}
\langle V(\mathbf{r}) \rangle = 0, \quad \langle V(\mathbf{r}) V(\mathbf{r'})  \rangle = \frac{V_0^2}{12} \delta(\mathbf{r}-\mathbf{r'}). 
\end{equation}
For each disorder realization $V(\mathbf{r})$ we then compute the retarded Green's function $G^R(\mathbf{r},\mathbf{r}',\omega)$ in the basis of onsite orbitals of the tight-binding model located at the points $\mathbf{r}$ and $\mathbf{r}'$:
\begin{equation}
G^R(\mathbf{r},\mathbf{r}',\omega) = (\omega + i 0^+ - \tilde{H}_0(\mathbf{r},\mathbf{r}') - U(\mathbf{r},\mathbf{r}'))^{-1}. 
\end{equation}
We can now reinstate the translational symmetry by averaging $G^R(\mathbf{r},\mathbf{r}',\omega)$ over 2000 disorder realizations:
\begin{equation}
G^R_{\mathrm{avg}}(\mathbf{r}-\mathbf{r}',\omega) = \langle G^R(\mathbf{r},\mathbf{r}',\omega) \rangle_\mathrm{dis}. 
\end{equation}
With the translational symmetry restored we can now express $G^R_{\mathrm{avg}}(\mathbf{r}-\mathbf{r}',\omega)$ in momentum space by performing Fourier transform
\begin{equation}
G^R(\mathbf{k}, \omega) = \int d(\mathbf{r}-\mathbf{r}') G^R_{\mathrm{avg}}(\mathbf{r}-\mathbf{r}',\omega) e^{i \mathbf{k}(\mathbf{r}-\mathbf{r}')}. 
\end{equation}
The final result of this procedure is the spectral function $A(\mathbf{k}, \omega)$ that is defined by 
\begin{equation}
\label{eq:spectral}
A(\mathbf{k}, \omega) = - \frac{1}{\pi} \mathrm{Im Tr} G^R.
\end{equation}

The other approach is to directly compute the self-energy using the integral equation obtained from the self-consistent Born approximation. We define the self-energy in the disordered system case as
\begin{equation}
(\omega - H_0 - \Sigma)^{-1} = \langle (\omega - H_0 - H_\mathrm{dis})^{-1} \rangle. 
\end{equation}
In the self-consistent Born approximation the self-energy $\Sigma_{\mathrm{SCBA}}(\omega)$ is independent of momentum and diagonal and is given by
\begin{equation}
\Sigma_{\mathrm{SCBA}}(\omega) = \frac{V_0^2 a^2}{12} \int_{\mathrm{BZ}} \frac{d^2k}{(2 \pi)^2} [\omega + i 0^+ - \tilde{H}_0(\mathbf{k}) - \Sigma_{\mathrm{SCBA}}(\omega)]^{-1}. 
\end{equation}
The equation is solved iteratively by numerical integration until the result converges. An example of the self-energy computed by using this method is shown in Fig. 3(a) of the main text. The self-energy obtained by this procedure is then used to compute the retarded Green's function directly in momentum space using $\tilde{H}_0(\mathbf{k})$:
\begin{equation}
\label{eq:retarded_gf}
G^R_{\mathrm{SCBA}}(\mathbf{k}, \omega) = (\omega + i 0^+ - \tilde{H}_0(\mathbf{k}) - \Sigma_{\mathrm{SCBA}}(\omega))^{-1}. 
\end{equation}
$G^R_{\mathrm{SCBA}}(\mathbf{k}, \omega)$ is finally used to calculate the spectral function using Eq.~(\ref{eq:spectral}).

\begin{figure}
\centering
\includegraphics[width=0.4\hsize]{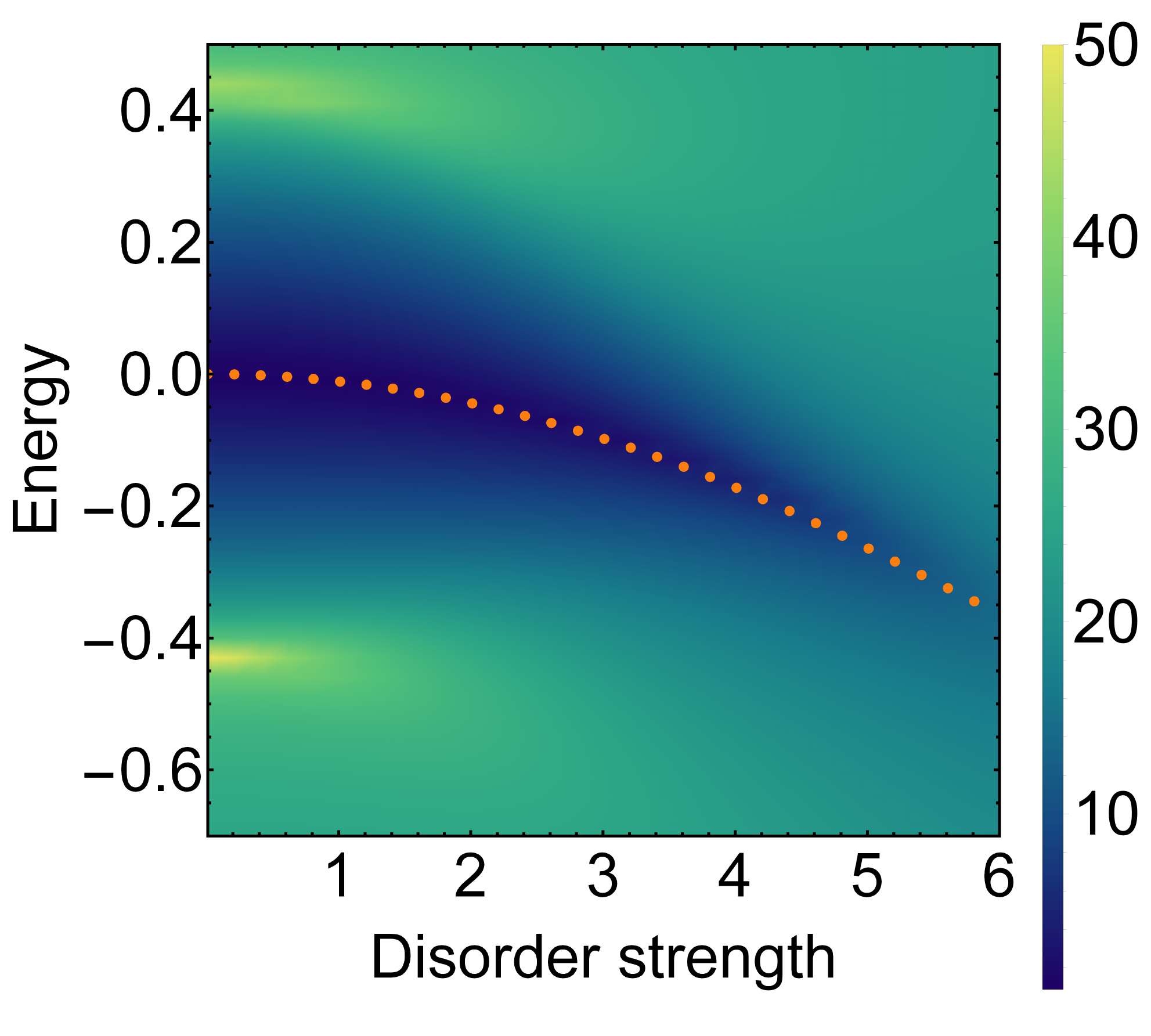}
\caption{Total integrated density of states $\rho(\omega)$ dependence on the disorder strength. The position of the minimum, which shifts to lower energies correspond to the position of the Dirac points in the case of infinite lifetimes. The points indicate the solutions of Eq.~(\ref{eq:omega_cond}).}
\label{fig:total_dos}
\end{figure}

In order to visualize the nodal arcs in the spectral function we can take a slice of $A(\mathbf{k}, \omega)$ across the whole Brillouin zone for a particular value of $\omega$. The Fermi surface will be an open contour for energy values corresponding to the position of Dirac point in the case of infinite lifetime. However, because the scattering generates both real and imaginary parts of self-energy and there is different disorder strength for both bands, the value of $\omega$ that results in a nodal arc is shifted from the position of the Dirac point in the clean system. Therefore, one has to determine by how much the disordered band is shifted and in this way find the correct value of energy to plot the spectral function at. If a term $\mathrm{Re} \Sigma_{11}$ is added to the Hamiltonian, the new position of the Dirac point will be $\mathrm{Re} \Sigma_{11}/2$. However, the self-energy is itself dependent on $\omega$, so in order to properly calculate the energy we have to solve
\begin{equation}
\label{eq:omega_cond}
\frac{\mathrm{Re} \Sigma(\omega)}{2} = \omega. 
\end{equation}
In Fig.~\ref{fig:total_dos} we plot the total integrated density of states $\rho(\omega)$
\begin{equation}
\rho(\omega) = \int_{BZ} d^2k A(\mathbf{k}, \omega). 
\end{equation}
as a function of the disorder strength $V_0$. As the disorder strength is increased, $|\mathrm{Re} \Sigma(\omega)|$ also becomes larger and the position of the minimum in the density of states moves into lower energies since $\mathrm{Re} \Sigma(\omega)<0$. The points indicate the solutions of Eq.~(\ref{eq:omega_cond}) and correctly describe the dependence of the energy shift of the minimum as the disorder strength changes.

\begin{figure}
\centering
\includegraphics[width=\hsize]{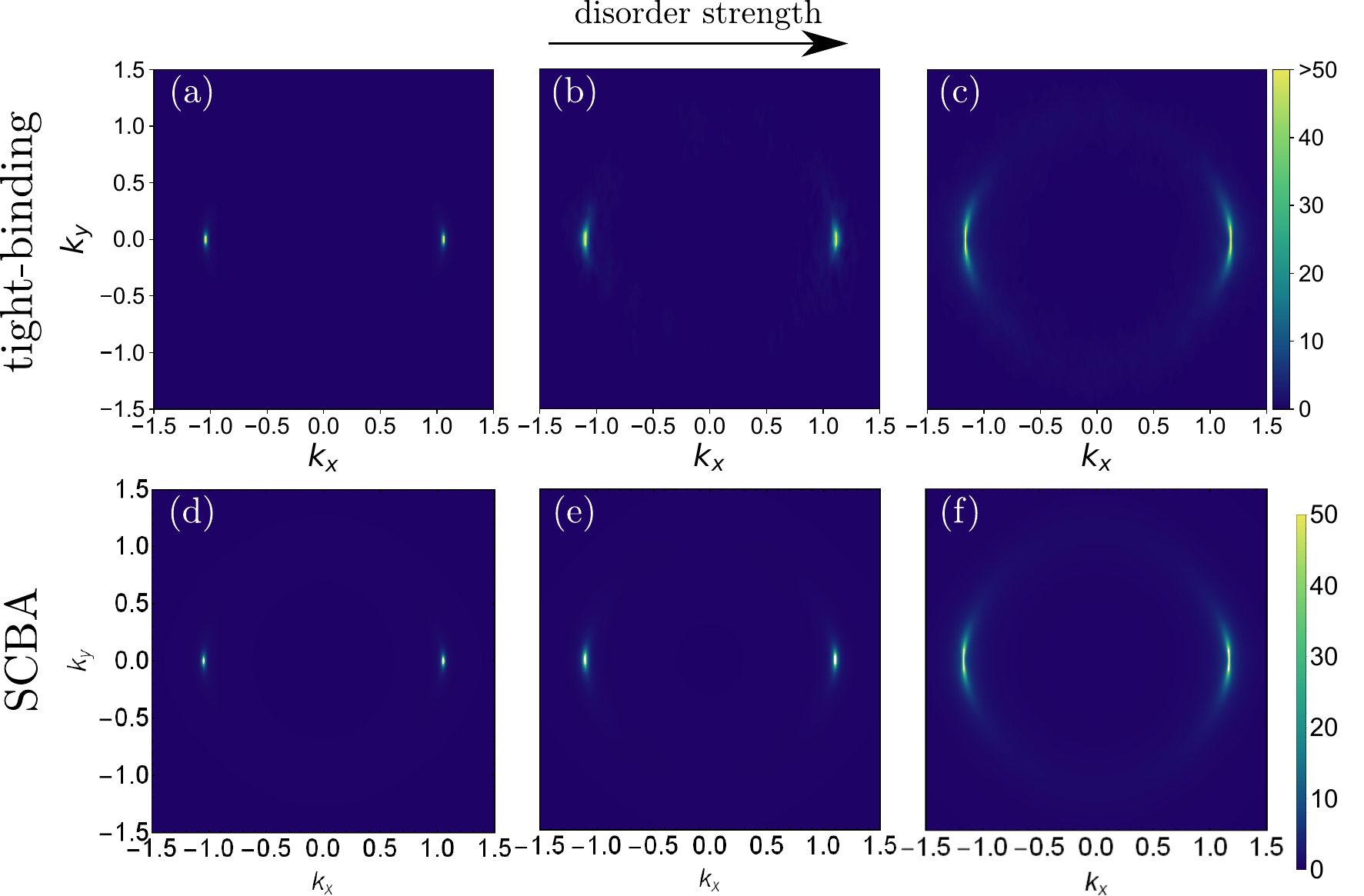}
\caption{Comparison of spectral function $A(\mathbf{k}, \omega)$ slices across a fragment of Brillouin zone for $\omega$ corresponding to the open contour Fermi surface, obtained using tight-binding lattice simulation (upper row) and self-consistent Born approximation (lower row). For each pair of upper and lower figure, the parameters used in the calculation are the same: $t=1$, $t_y=0.25$, $u=3.0$ and disorder strengths $V_0 = 0.5, 3.0, 4.5$, respectively. The spectral functions show an excellent agreement between both methods.}
\label{fig:TB_SCBA_supplement}
\end{figure}

\begin{figure}
\includegraphics[width=0.8\hsize]{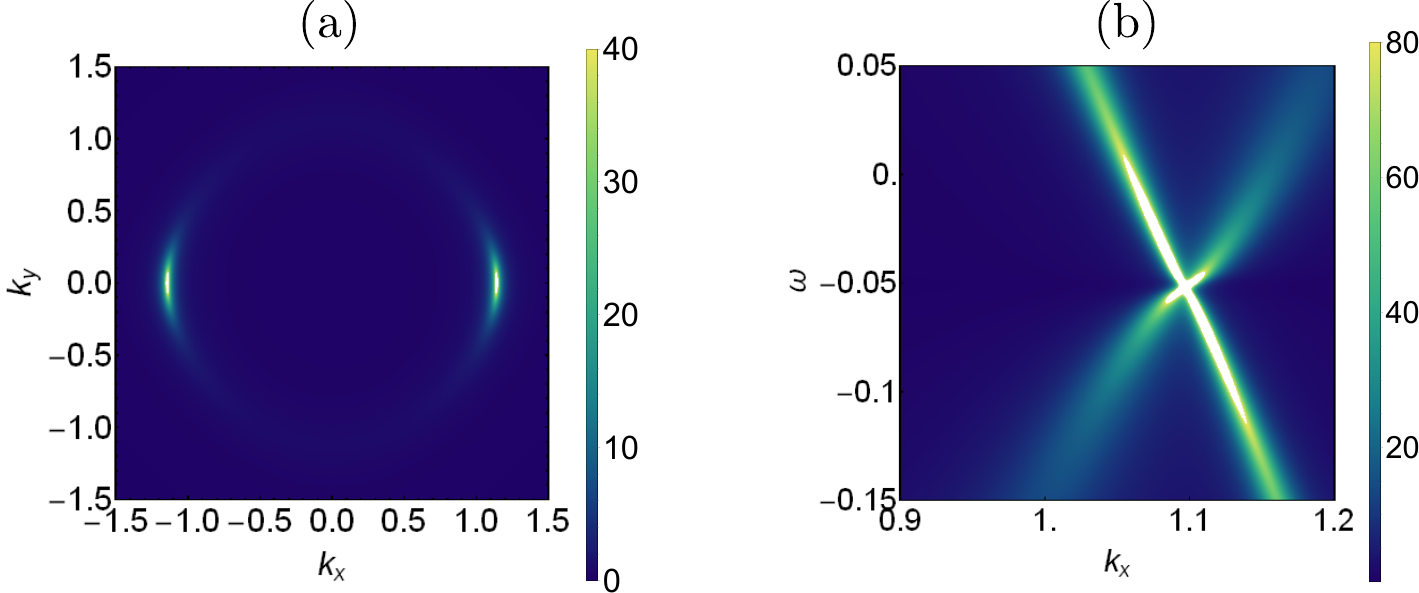}
\caption{(a) Spectral function $A(\mathbf{k}, \omega)$ slices across the whole Brillouin zone for $\omega$ corresponding to the open contour Fermi surface, obtained using self-consistent Born approximation with disorder on both bands ($V_1=3.5$ on the first band $V_2=1.75$ on the second band). (b) Spectral function $A(\bm{k}, \omega)$ slices for $k_y=0$ with disorder on both bands ($V_1=2.5$ on the first band $V_2=1.25$ on the second band). The model parameters used in both calculation are the same: $t=1$, $t_y=0.2$, $u=3.0$ (different from those in the main text).}
\label{fig:disorder_on_both}
\end{figure}

With the correct value of $\omega$ that gives an open contour for the Fermi surface we can now analyze the behavior of the spectral function. In Fig.~\ref{fig:TB_SCBA_supplement} we present a comparison of the spectral functions obtained using both methods described above for increasing values of disorder strength. In each pair, the same parameters have been used for both tight-binding and SCBA calculation: $t=1$, $t_y=0.25$, $u=3.0$ and disorder strengths $V_0 = 0.5, 3.0, 4.5$, respectively. Spectral functions obtained with both methods show an excellent agreement and confirm that our system is well described by the SCBA. This allows us to use this method in all the cases that require high energy resolution in order to minimize the computational costs.

In Fig.~\ref{fig:disorder_on_both} we show that our conclusions remain virtually the same even if disorder is included in both orbitals and parameters of the model are changed. In both panels the ratio of disorder strengths on both orbitals is 2:1 and hopping parameters are changed from the main text (now they are $t=1$, $t_y=0.2$, $u=3.0$). Fig.~\ref{fig:disorder_on_both}(a) show a clear nodal arc, which is broadened compared to the arc presented in the main text due to the disorder on both bands. Fig.~\ref{fig:disorder_on_both}(b) shows that the Dirac cone still tilts significantly, while both bands are broadened now. This shows that our results are generic and rely only on the existence of asymmetry in scattering of both components of the Dirac fermion.

\begin{figure}
\centering
\includegraphics[width=0.36\hsize]{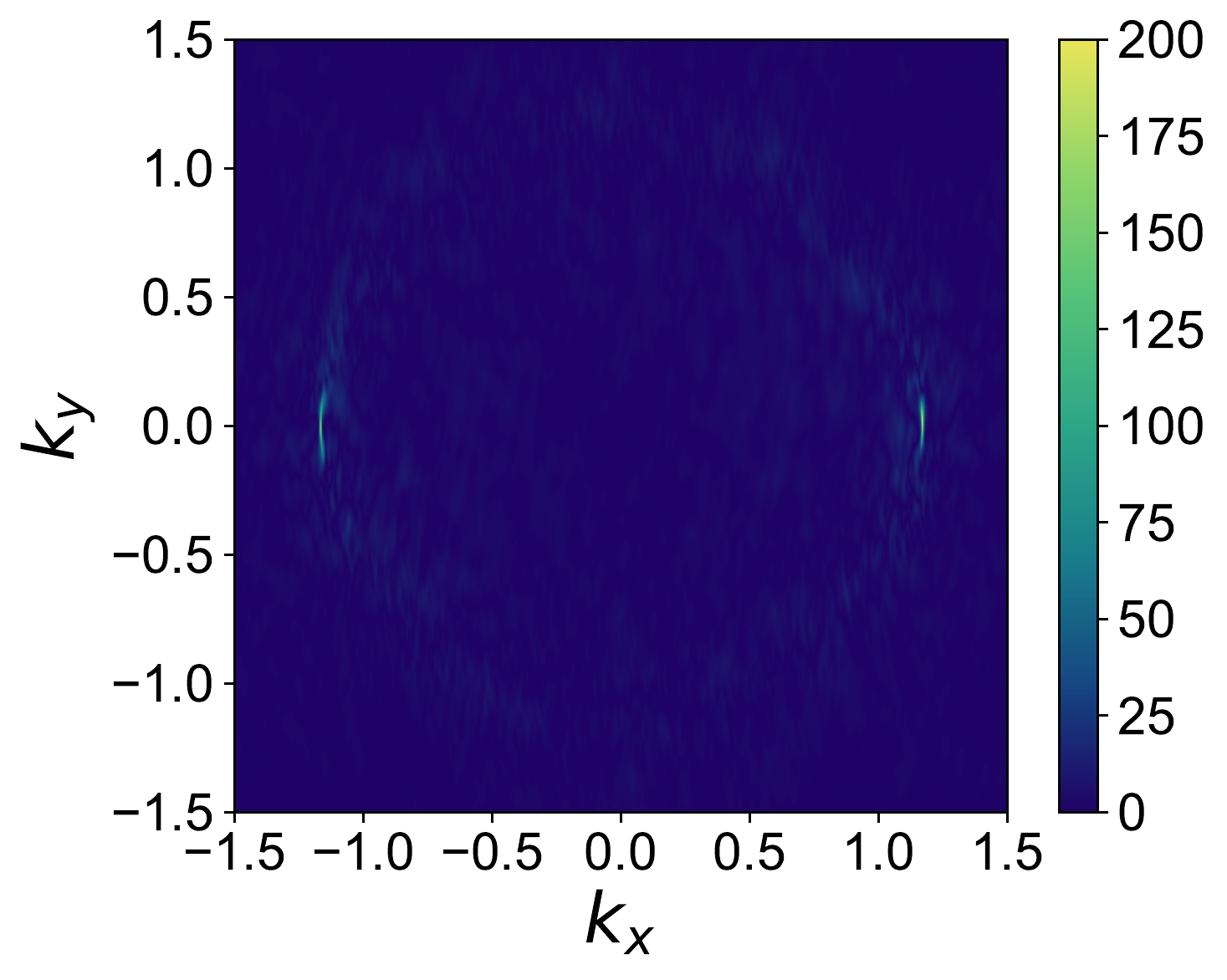}
\caption{Spectral function $A(\mathbf{k}, \omega)$ computed as a discrete Fourier transform of the real space Green's function without disorder averaging (single disorder realization) for a lattice with $3500 \times 3500$ sites. The nodal arc is clearly visible even without averaging procedure and full restoration of translational invariance. The parameters used in the calculation are: $t=1$, $t_y=0.25$, $u=3.0$ and disorder strength $V_0 = 0.5, 3.0, 4.5$.}
\label{fig:no_averaging}
\end{figure}

In Fig.~\ref{fig:no_averaging} we show the spectral function obtained as a discrete Fourier transform of a calculation for a single disorder realization. Even though no averaging is performed and translational invariance is not fully restored, the spectral function shows a nodal arc at the same coordinates as in the averaged case. This suggests that the arc we observe is a feature of a particular disordered sample and not just a result present only in a virtual crystal with average disorder potential.

\begin{figure}
\centering
\includegraphics[width=0.65\hsize]{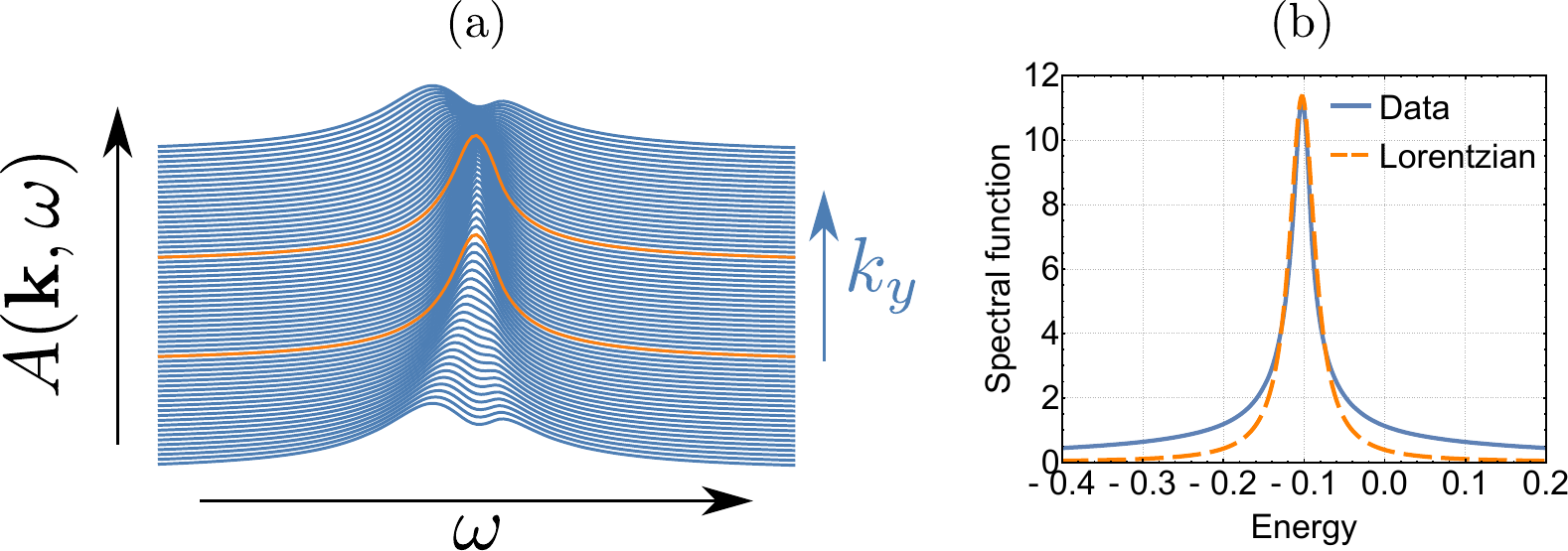}
\caption{(a) Spectral function $A(\omega)$ for several values of $k_y$ calculated for $k_x = 1.143$ with disorder present on both orbitals. The orange lines highlight the exceptional point at $k_y \approx 0.04$. (b) Comparison of the spectral function close to the exceptional point to a Lorentzian function with area renormalized by reduced quasiparticle weights $Z_{11}$ and $Z_{22}$.}
\label{fig:exceptional_point}
\end{figure}

In Fig.~\ref{fig:exceptional_point}(a) we show several line cuts of spectral function $A(\omega)$ for different $k_y$ calculated for $k_x=1.143$. The orange lines highlight the curves for the exceptional point, where both complex poles of the retarded Green's function from Eq.~\eqref{eq:retarded_gf} coalesce and are approximately described by a Lorentzian curve with the width determined by the lifetimes of both orbitals. Fig.~\ref{fig:exceptional_point}(b) shows a comparison between a Lorentzian function with area under the curve determined by sum of quasiparticle weights $Z_{11} + Z_{22}$ and the spectral function for $\mathbf{k}$ close to the exceptional point. The agreement is very good for energies close to the nodal arc, where the approximation of energy-independent lifetime $\Gamma_0$ holds.


\begin{thebibliography}{99}

\bibitem{Fradkin}
E. Fradkin, Phys. Rev. B \textbf{33}, 3257 (1986); \textbf{33}, 3263 (1986).

\bibitem{Guruswamy}
S. Guruswamy, A. LeClair, and A. W. W. Ludwig, Nucl. Phys. B \textbf{583}, 475 (2000).

\bibitem{review}
For a review, see e.g., F. Evers and A. D. Mirlin, Rev. Mod. Phys. \textbf{80}, 1355 (2008).

\bibitem{Ludwig}
A. W. W. Ludwig, M. P. A. Fisher, R. Shankar, and G. Grinstein, Phys. Rev. B \textbf{50}, 7526 (1994).

\bibitem{Nersesyan}
A. A. Nersesyan, A. M. Tsvelik, and F. Wenger, Nucl. Phys. B \textbf{438}, 561 (1995).

\bibitem{SenthilFisher}
T. Senthil and M. P. A. Fisher, Phys. Rev. B \textbf{61}, 9690 (2000).

\bibitem{Bocquet}
M. Bocquet, D. Serban, and M. R. Zirnbauer, Nucl. Phys. B \textbf{578}, 628 (2000).

\bibitem{Altland}
A. Altland, B. D. Simons, and M. R. Zirnbauer, Phys. Rep. \textbf{359}, 283 (2002).

\bibitem{Foster}
Matthew S. Foster, Hong-Yi Xie, and Yang-Zhi Chou, Phys. Rev. B \textbf{89}, 155140 (2014).

\bibitem{McCann}
E. McCann, K. Kechedzhi, V. I. Fal’ko, H. Suzuura, T. Ando, and B. L. Altshuler, Phys. Rev. Lett. \textbf{97}, 146805 (2006).

\bibitem{Koshino}
M. Koshino and T. Ando, Phys. Rev. B \textbf{73}, 245403 (2006).

\bibitem{Ostrovsky}
P. M. Ostrovsky, I. V. Gornyi, and A. D. Mirlin, Phys. Rev. B \textbf{74}, 235443 (2006).

\bibitem{Aleiner}
I. L. Aleiner and K. B. Efetov, Phys. Rev. Lett. \textbf{97}, 236801 (2006).

\bibitem{Altland2}
A. Altland, Phys. Rev. Lett. \textbf{97}, 236802 (2006).


\bibitem{Ostrovsky2}
P. M. Ostrovsky, I. V. Gornyi, and A. D. Mirlin, Phys. Rev. Lett. \textbf{98}, 256801 (2007).

\bibitem{Bardarson}
J. H. Bardarson, J. Tworzyd{\l}o, P. W. Brouwer, and C. W. J. Beenakker, Phys. Rev. Lett. \textbf{99}, 106801 (2007).

\bibitem{Nomura}
K. Nomura, M. Koshino, and S. Ryu, Phys. Rev. Lett. \textbf{99}, 146806 (2007).

\bibitem{Ryu2}
S. Ryu, C. Mudry, H. Obuse, and A. Furusaki, Phys. Rev. Lett. \textbf{99}, 116601 (2007).

\bibitem{Noro}
M. Noro, M. Koshino, and T. Ando, J. Phys. Soc. Jpn. \textbf{79}, 094713 (2010).

\bibitem{Mong}
R. S. K. Mong, J. H. Bardarson, and J. E. Moore, Phys. Rev. Lett. \textbf{108}, 076804 (2012).

\bibitem{FuKane}
L. Fu and C. L. Kane, Phys. Rev. Lett. \textbf{109}, 246605 (2012).

\bibitem{ZeljkovicMappingunconventionalorbital2014}
I. Zeljkovic, Y. Okada, C.-Y. Huang, R. Sankar, D. Walkup, W. Zhou, M. Serbyn, F. Chou, W.-F. Tsai, H. Lin, A. Bansil, L. Fu, M. Z. Hasan, and V. Madhavan,
Nat. Phys. \textbf{10}, 572–577 (2014)

\bibitem{KoziiNonHermitianTopologicalTheory2017}%
V. Kozii and L. Fu, arXiv:1708.05841.

\bibitem{Dai}
Y. Xu, C. Yue, H. Weng, and X. Dai, Phys. Rev. X \textbf{7}, 011027 (2017).

\bibitem{ShenTopologicalBandTheory2017}
H. Shen, B. Zhen, and L. Fu, arXiv:1706.07435.

\bibitem{WeylFlatBand}
A. A. Zyuzin and A. Yu. Zyuzin, Phys. Rev. B \textbf{97}, 041203(R) (2018)

\bibitem{Ando}
Y. Tanaka, Zhi Ren, T. Sato, K. Nakayama, S. Souma, T. Takahashi, Kouji Segawa and  Yoichi Ando, Nature Physics, \textbf{8}, 800 (2012). 

\bibitem{SodemannFu}
Inti Sodemann and Liang Fu, Phys. Rev. Lett. \textbf{115}, 216806 (2015). 

\bibitem{tilted1}
A. Kobayashi, S. Katayama, Y. Suzumura and H. Fukuyama, J. Phys. Soc. Jpn. \textbf{76}, 034711 (2007); A. Kobayashi, Y. Suzumura, and H. Fukuyama, J. Phys. Soc. Jpn. \textbf{77}, 064718 (2008). 

\bibitem{tilted2}
M. O. Goerbig, J.-N. Fuchs, G. Montambaux, and F. Pi{\'e}chon,
Phys. Rev. B \textbf{78}, 045415 (2008).

\bibitem{Isobe}
H. Isobe and N. Nagaosa, J. Phys. Soc. Jpn. \textbf{81}, 113704 (2012);  Phys. Rev. Lett. \textbf{116}, 116803 (2016).

\bibitem{general}
H. Isobe, M. Papaj, and L. Fu (unpublished).

\bibitem{MUMPS}
P. R. Amestoy, I. S. Duff, J. S. Koster, J. Y. L’Excellent
SIAM. J. Matrix Anal. Appl. \textbf{23} (1), 15 (2001)

\bibitem{KWANT}
C. W. Groth, M. Wimmer, A. R. Akhmerov, X. Waintal, New J. Phys. \textbf{16}, 063065 (2014).

\bibitem{SM}
See the Supplemental Material for the details of the numerical simulations.


\end{thebibliography}
\end{document}